\documentclass[a4paper,conference,compressed,twocolumn]{IEEEtran}

\usepackage[pdftex]{graphicx}
\usepackage{color,colortbl,multicol,amssymb,amsfonts,array}
\usepackage[cmex10]{amsmath}
\usepackage{gensymb,textcomp}
\usepackage[ansinew,utf8]{inputenc}
\usepackage[acronym,shortcuts,nomain,nopostdot]{glossaries}

\definecolor{hyp}{rgb}{0,0,0.3} 
\usepackage[colorlinks=true,linkcolor=hyp,filecolor=hyp,citecolor=hyp,urlcolor=hyp]{hyperref}

\newglossarystyle{acro}{
    \glossarystyle{super}

}
\newcommand{\acro}[2]{\newacronym{#1}{#1}{#2}}      
\newcommand{\acrol}[3]{\newacronym{#1}{#2}{#3}}     

\newcommand{\acrop}[4]{\newacronym[plural=#3,firstplural=#4 (#3)]{#1}{#1}{#2}}

\newcommand{\acF}[1]{\acf{#1}\glsunset{#1}}

\newcommand{\acXp}[1]{\acrlongpl{#1}}

\acro{3GPP}{3rd generation partnership project}
\acrol{1-D}{\mbox{1-D}}{one-dimensional}
\acrol{2-D}{\mbox{2-D}}{two-dimensional}
\acrol{3-D}{\mbox{3-D}}{three-dimensional}
\acrol{4-D}{\mbox{4-D}}{four-dimensional}
\acrol{6-D}{\mbox{6-D}}{six-dimensional}
\acro{3G}{third generation}
\acro{4G}{fourth generation}
\acro{5G}{fifth generation}

\acrop{AoA}{azimuth angle of arrival}{AoAs}{azimuth angles of arrival}
\acrop{AoD}{azimuth angle of departure}{AoDs}{azimuth angles of departure}
\acro{ASA}{azimuth spread of arrival}
\acro{AS}{angular spread}
\acro{ASD}{azimuth spread of departure}
\acro{AWGN}{additive white Gaussian noise}
\acro{AGC}{adaptive gain control}
\acro{ACF}{autocorrelation function}
\acro{ASE}{average squared error}

\acro{BD}{block diagonalization}
\acro{BS}{base station}
\acro{BC}{broadcast channel}
\acro{BP}{break point}

\acro{CDF}{cumulative distribution function}
\acro{CIR}{channel impulse response}
\acro{CoMP}{coordinated multi-point}
\acro{CoF}{cost of feedback}
\acro{CRONOS}{cellular radio network simulator}
\acro{CSI}{channel state information}
\acro{COST}{European Cooperation in Science and Technology}

\acro{DS}{delay spread}
\acro{DD}{Doppler-delay}
\acro{DFT}{discrete Fourier transform}
\acro{DPC}{dirty-paper coding}
\acro{D2D}{device-to-device}

\acro{ESA}{elevation angle spread of arrival}
\acro{ESD}{elevation angle spread of departure}
\acrop{EoA}{elevation angle of arrival}{EoAs}{elevation angles of arrival}
\acrop{EoD}{elevation angle of departure}{EoDs}{elevation angles of departure}

\acro{FBR}{feedback rate}
\acro{FBS}{first-bounce scatterer}
\acro{FI}{feedback interval}
\acro{FR}{frequency response}
\acro{FR1}{frequency reuse 1}
\acro{FR4}{frequency reuse 4}
\acro{FIR}{finite impulse response}
\acro{FFT}{fast Fourier transform}
\acro{FWHM}{full width at half maximum}
\acro{FDD}{frequency-division duplexing}

\acro{GCS}{global coordinate system}
\acro{GF}{geometry factor}
\acro{GR}{ground reflection}
\acro{GoC}{gain of CoMP}
\acro{GoP}{gain of prediction}
\acro{GPS}{global positioning system}
\acro{GSCM}{geometry-based stochastic channel model}

\acro{HHI}{Heinrich Hertz Institute}

\acro{ISD}{inter site distance}
\acrol{iid}{i.i.d.}{independent and identically distributed}
\acro{IDFT}{inverse discrete Fourier transform}
\acro{IFFT}{inverse fast Fourier transform}
\acro{IR}{impulse response}
\acro{ITU}{International Telecommunication Union}

\acro{JT}{joint transmission}
\acro{JCF}{joint correlation function}

\acro{KF}{Ricean K-factor}
\acro{KPI}{key performance indicator}

\acro{LBS}{last-bounce scatterer}
\acro{LHCP}{left hand circular polarized}
\acro{LOS}{line of sight}
\acro{LSP}{large-scale parameter}
\acro{LTE}{long term evolution}
\acro{LSAS}{large-scale antenna systems}
\acro{LSF}{large-scale fading}

\acro{MET}{multi-user eigenmode transmission}
\acro{MIMO}{multiple-input multiple-output}
\acro{MISO}{multiple-input single-output}
\acro{MMSE}{minimum mean square error}
\acro{MPC}{multipath component}
\acro{MS}{mobile station}
\acro{MSE}{mean square error}
\acro{MT}{mobile terminal}
\acro{MU-MIMO}{multi-user MIMO}
\acro{MIMOSA}{MIMO over satellite}
\acro{MAC}{multiple-access channel}
\acro{MRT}{maximum ratio transmission}

\acro{NLOS}{non-line of sight}

\acro{OFDM}{orthogonal frequency division multiplexing}
\acro{O2I}{outdoor-to-indoor}

\acro{PDP}{power delay profile}
\acro{PDF}{probability density function}
\acro{PG}{path gain}
\acro{PL}{path loss}
\acro{PUCA}{polarized uniform circular array}
\acro{PAS}{power-angular spectrum}
\acro{P2P}{peer-to-peer}

\acro{QAM}{quadrature amplitude modulation}
\acro{QuaDRiGa}{quasi deterministic radio channel generator}
\acro{QRDRLS}{QR-decomposition recursive least squares}

\acro{RB}{resource block}
\acro{RHCP}{right hand circular polarized}
\acro{RLS}{recursive least-squares}
\acro{RLE}{run-length encoding}
\acro{Rx}{receiver}
\acro{RMS}{root mean square}
\acro{RAN}{radio access network}
\acro{RSRP}{reference signal received power}

\acro{SAGE}{space-alternating generalized expectation-maximization}
\acro{SCM}{spatial channel model}
\acro{SF}{shadow fading}
\acro{SV}{singular value}
\acro{SVD}{singular value decomposition}
\acro{SNR}{signal to noise ratio}
\acro{SIR}{signal to interference ratio}
\acro{SINR}{signal to interference and noise ratio}
\acro{SISO}{single input single output}
\acro{SIMO}{single input multiple output}
\acro{STD}{standard deviation}
\acro{SSG}{state sequence generator}
\acro{SSF}{small-scale-fading}
\acro{SOS}{sum-of-sinusoids}

\acro{TP}{throughput}
\acro{TUB}{Technische Universit\"{a}t Berlin}
\acro{Tx}{transmitter}

\acro{ULA}{uniform linear array}
\acro{UML}{unified modeling language}
\acro{UMi}{urban-microcell}
\acro{UMa}{urban-macrocell}
\acro{UAV}{unmanned aerial vehicle}

\acro{VR}{visibility region}

\acro{WINNER}{Wireless World Initiative for New Radio}
\acro{WSSUS}{wide sense stationary uncorrelated scattering}
\acro{WSS}{wide-sense stationary}
\acro{WGS}{world geodetic system}

\acro{XPD}{cross-polarization discrimination}
\acro{XPR}{cross polarization ratio}

\acro{ZF}{zero-forcing}
\acro{ZoA}{zenith angle of arrival}
\acro{ZoD}{zenith angle of departure}
\acro{ZSA}{zenith angle spread of arrival}
\acro{ZSD}{zenith angle spread of departure}

\newcolumntype{L}[1]{>{\raggedright\let\newline\\\arraybackslash\hspace{0pt}}m{#1}}
\newcolumntype{C}[1]{>{\centering\let\newline\\\arraybackslash\hspace{0pt}}m{#1}}
\newcolumntype{R}[1]{>{\raggedleft\let\newline\\\arraybackslash\hspace{0pt}}m{#1}}

\begin{document}

\author{\IEEEauthorblockN{ Stephan Jaeckel\IEEEauthorrefmark{1}, Leszek Raschkowski\IEEEauthorrefmark{1}, Frank Burkhardt\IEEEauthorrefmark{2} and Lars Thiele\IEEEauthorrefmark{1}}
\IEEEauthorblockA{\IEEEauthorrefmark{1} Fraunhofer Heinrich Hertz Institute, Berlin, Germany, stephan.jaeckel@hhi.fraunhofer.de}\IEEEauthorblockA{\IEEEauthorrefmark{2} Fraunhofer Institute for Integrated Circuits, Erlangen, Germany}}

\title{Efficient Sum-of-Sinusoids based Spatial Consistency for the 3GPP New-Radio Channel Model}

\maketitle

\begin{abstract}
Spatial consistency was proposed in the 3GPP TR 38.901 channel model to ensure that closely spaced mobile terminals have similar channels. Future extensions of this model might incorporate mobility at both ends of the link. This requires that all random variables in the model must be correlated in 3 (single-mobility) and up to 6 spatial dimensions (dual-mobility). Existing filtering methods cannot be used due to the large requirements of memory and computing time. The sum-of-sinusoids model promises to be an efficient solution. To use it in the 3GPP channel model, we extended the existing model to a higher number of spatial dimensions and propose a new method to calculate the sinusoid coefficients in order to control the shape of the autocorrelation function. The proposed method shows good results for 2, 3, and 6 dimensions and achieves a four times better approximation accuracy compared to the existing model. This provides a very efficient implementation of the 3GPP proposal and enables the simulation of many communication scenarios that were thought to be impossible to realize with geometry-based stochastic channel models.
\end{abstract}

\vspace{\baselineskip}


\section{Introduction}


Channel models are an essential requirement for the development of new wireless communication protocols and systems. Standardized channel models can be used to evaluate and compare the different proposals against each other. For the upcoming \acF{5G} of wireless cellular communications, a new channel model was proposed by the \ac{3GPP} \cite{3gpp_tr_38901_v1410}. An open-source implementation of this model is provided by the \ac{QuaDRiGa} \cite{Jaeckel2017_diss,quadriga_www}. An essential new feature that was introduced by \ac{3GPP} is \emph{spatial consistency} which solves one major drawback of previous \acp{GSCM}, namely the lack of realistic correlation in the \acF{SSF}. Without spatial consistency, the positions of individual scattering clusters are generated randomly for each new \ac{MT} position. This contradicts the causality and the observations made in channel measurements \cite{Narandzic2011}.

\Acp{GSCM} consist of two main components: a stochastic part that generates a random propagation environment, and a deterministic part that lets transmitters and receivers interact with this environment. In order to make realistic predictions of the wireless system performance, the random environment must fulfill certain statistical properties which are determined by measurements. This means that for a given set of model parameters the joint spatial correlation of these parameters must be captured for a large number of transceivers. This is done in the so-called \ac{LSF} model. A subsequent \ac{SSF} model then generates individual scattering clusters for each \ac{MT}.


\Acp{LSP} are more or less constant within an area of several meters. An example for this is the \ac{SF} which is caused by buildings or trees blocking a significant part of the signal. The so-called decorrelation distance of the \ac{SF}, \emph{i.e.}, the distance a \ac{MT} must move to experience a significant change in the \ac{SF}, is in the same order of magnitude as the size of the objects causing it. Thus, if a \ac{MT} travels along a trajectory or if multiple \acp{MT} are closely spaced together, their \acp{LSP} are correlated. A common approach to model such correlation is by filtered Gaussian-distributed random numbers \cite{Bakowski2011}. However, when it comes to spatial consistency, the positions of the scattering clusters must also be spatially correlated. The \ac{3GPP} proposal suggests that ``spatially consistent powers/delays/angles of clusters are generated" \cite{3gpp_tr_38901_v1410}. However, this requires that all random variables that determine the location of the scattering clusters are correlated. For a moderate scenario with 12 clusters and 20 sub-paths per cluster, this results in 2288 random variables\footnote{In the 3GPP new-radio model, scattering is based on the LOS / NLOS state (1 variable); \acp{LSP} (7 variables: delay spread, shadow fading, 4 angular spreads, K-factor); delays, powers, per-cluster angles ($10\cdot 12$ variables); random coupling of sub-paths ($4 \cdot 12 \cdot 20$ variables); Cross polarization power ratios ($12 \cdot 20$ variables); Initial random phases ($4 \cdot 12 \cdot 20$ variables) }. Compared to the 7 variables needed for the \ac{LSF} model, the filtering approach therefore requires a prohibitively large amounts of memory and computing time.

Another problem arises when incorporating so-called \emph{vertical industries} into the \ac{5G} infrastructure which is not yet covered by the 3GPP new-radio model but discussed in several ongoing research projects and standardization activities. Such \emph{verticals} could be vehicular networks, air-to-ground communications, industrial \ac{P2P} communications, or communication scenarios involving satellites in low-earth orbit. All of these examples have in common that both ends of the link are mobile. However, the simultaneous mobility of both communication partners is not supported by the classical cellular shadowing models. Therefore, an alternative method for the generation of the random propagation environment is needed \cite{Jamsa2015}.


A computational efficient method to generate correlated random variables has been introduced by P\"{a}tzold el.\ al.\ \cite{Patzold1996} who approximated the filtered white Gaussian noise process by a finite sum of properly weighted sinusoids. This idea was further developed into a \ac{2-D} shadowing model \cite{Cai2003}. Wang et.\ al.\ then extended this \ac{2-D} method into a \ac{4-D} method for \ac{D2D} channels \cite{Wang2005b}. A common problem for all these methods is finding the coefficients that best approximate a desired \ac{ACF} with a limited number of sinusoids.


In this paper, a new approximation method is presented that allows the efficient calculation of the sinusoid coefficients for an arbitrary \ac{ACF} in \ac{3-D} space. Previously published results focused solely on the exponential \ac{ACF} and mobility was limited to planar movements. Furthermore, it is shown how this translates directly into a six-dimensional random process for \ac{D2D} channels where both ends can be in different propagation environments, such as in air-to-ground channels. Compared to \cite{Wang2005b}, the proposed method also achieves a 6.5~dB better approximation \ac{ASE} with the same number of sinusoids. An implementation of this model is available in MATLAB and Octave as part of the open-source \ac{QuaDRiGa} channel model \cite{quadriga_www}.

\section{The Sum of Sinusoids Model}

A spatially correlated Gaussian random process generates Normal distributed random numbers
\begin{equation}
    k(x,y,z) \sim \mathcal{N}(0,1)\text{,}
\end{equation}
with zero mean and unit variance. The value $k$ is a function of the terminal location in \ac{3-D} Cartesian coordinates $(x,y,z)$. From these numbers, other types of distributions can be obtained, e.g.\ Uniform or log-Normal distributed random numbers having a different mean or variance. The \ac{1-D} spatial \acF{ACF} describes how fast the local mean of $k(x,y,z)$ evolves as a terminal moves. The \ac{ACF} is usually modeled as an exponential decay function
\begin{equation}\label{eq:exponential_corr_decay_map}
    \rho(d) = \exp\left(-\frac{d}{d_\lambda}\right)\text{,}
\end{equation}
with $d$ as the distance between two positions and $d_\lambda$ as the so-called decorrelation distance, i.e.\ the distance at which the correlation between two samples falls below $e^{-1}\approx 0.37$ \cite{Gudmundson1991_Correlation}. However, other types of decay functions may be desirable. The \acF{SOS} method outlined in \cite{Patzold1996} approximates a \ac{1-D} Gaussian random process $k(x)$ as a function of the position $x$ on a \ac{1-D} linear trajectory as
\begin{equation}\label{eq:sos_1D_process}
    \hat{k}(x) = \sum_{n=1}^N a_n  \cos\left\{ 2\pi f_n x + \psi_n \right\}
\end{equation}
with $N$ sinusoids. The variables $a_n$, $f_n$, and $\psi_n$ denote the amplitude, the frequency, and the phase of a sinusoid, respectively. The amplitudes $a_n$ and the frequencies $f_n$ are determined in a way that $\hat{k}(x)$ has similar statistical properties as $k(x)$, i.e.\ $\hat{k}(x)$ has the same approximate \ac{ACF} and the \ac{CDF} is close to Gaussian density if $N$ is sufficiently large. According to \cite{Patzold1996}, 6 to 30 sinusoids are sufficient for a \ac{1-D} approximation. The phases $\psi_n$ are random variables distributed in the range from $-\pi$ to $\pi$. Hence, exchanging the $\psi_n$ while keeping $a_n$ and $f_n$ fixed creates a new set of spatially correlated random variables at minimal computational cost. A straight-forward expansion to a \ac{3-D} Gaussian random process follows from \cite{Wang2005b} as
\begin{equation}\label{eq:sos_3D_process}
    \hat{k}(x,y,z) = \sum_{n=1}^N a_n  \cos\left\{ 2\pi \left(f_{x,n} x + f_{y,n} y + f_{z,n} z\right) + \psi_n \right\}\text{.}
\end{equation}
Under the assumption that the fluctuations of $k(x,y,z)$ are wide sense stationary, the \ac{ACF} only depends on the distance between two terminal positions. Hence, the \ac{3-D} spatial \ac{ACF} of $\hat{k}(x,y,z)$ can be expressed as \cite{Wang2005b}
\begin{multline}\label{eq:sos_approximate_acf}
    \hat{\rho}(\Delta x,\Delta y,\Delta z) = \\
    \sum_{n=1}^N \frac{a_n^2}{2} \cos\left\{ 2\pi \left(f_{x,n} \Delta x + f_{y,n} \Delta y + f_{z,n} \Delta z\right) \right\}\text{.}
\end{multline}
When a \ac{MT} moves from one location to another, the correlation $\rho(\Delta x,\Delta y,\Delta z)$ between the generated values $k(x,y,z)$ depends not only on the distance, but also on the direction of movement. In \ac{3-D} space, the traveling direction can be expressed by pitch and yaw. The pitch angle $\theta$ describes the vertical (tilt) angle relative to the horizontal plane. Positive rotation is up. The bearing or yaw angle $\phi$ describes the orientation on the ground. Here, it is defined in mathematic sense, i.e., seen from above, a value of 0 points to the east and the angles increase counter-clockwise. By assigning a direction $(\theta_n,\phi_n)$ to each of the $N$ sinusoids, the relative movement in $x$, $y$, and $z$ direction can be expressed as
\begin{eqnarray}
\label{eq:sos_x_movement} f_{x,n} \Delta x &=& f_n \cdot d \cdot \cos \phi_n  \cos \theta_n, \\
\label{eq:sos_y_movement} f_{y,n} \Delta y &=& f_n \cdot d \cdot \sin \phi_n  \cos \theta_n, \\
\label{eq:sos_z_movement} f_{z,n} \Delta z &=& f_n \cdot d \cdot \sin \theta_n,
\end{eqnarray}
where $f_n$ is the root-frequency of the $n$-th sinusoid frequency and $d$ is the distance. The directions $(\theta_n,\phi_n)$ have to be chosen in a way that covers all possible movement directions of the \ac{MT}. This can be done by generating equidistributed points on the surface of a sphere as described in \cite{Deserno2004}.

P\"{a}tzold et.\ al.\ \cite{Patzold1996} proposed four methods to determine the amplitudes $a_n$ and frequencies $f_n$ in a \ac{SOS} model. All of them show different performance in terms of \acF{ASE} vs.\ the number of sinusoids, and computational complexity. They have in common that the approximated \ac{ACF} is an exponential decay function \eqref{eq:exponential_corr_decay_map}. However, this is not always desirable since other types of decay functions might be needed. Hence, in the next section, a numeric approximation method is presented that calculates the sinusoid coefficients for arbitrary \acp{ACF}.

\section{Approximation of Arbitrary ACFs}\label{sec:sos_arbitrary_approx}

The approximation method is derived from the Monte Carlo method \cite{Hoeher1992,Patzold1996}. The sinusoid frequencies $f_n$ are generated by an iterative optimization method that minimizes the error between the desired \ac{ACF} and the approximate \ac{ACF} for a given number of sinusoids. This method requires that the \ac{ACF} is discretely sampled at $s = 1 \ldots S$ sampling distances. This is done by defining a vector $\mathbf{d}$ that contains the sampling distances in increasing order.
\begin{equation}\label{eq:sos_dist_vec}
    \mathbf{d} =
    \left(
      \begin{array}{cccc}
        d_1 & d_2 & \ldots & d_S \\
      \end{array}
    \right)^T
\end{equation}
Then, the sampled \ac{ACF} $\rho(\mathbf{d})$ is obtained. The first distance value $d_1$ must be 0 and the first correlation value $\rho_1$ must be 1, i.e.\ at zero-distance the generated values $\hat{k}(x,y,z)$ are identical. The $N$ root-frequencies are randomly initialized to
\begin{equation}
    f_n \sim \frac{1}{d_S} \cdot \mathcal{U}(-\pi,\pi)\text{,}
\end{equation}
where $\mathcal{U}(-\pi,\pi)$ describes an uniform distribution with values between $-\pi$ and $\pi$, and $d_S$ the maximum distance for which the \ac{ACF} is defined. The directional components $f_{x,n}$, $f_{y,n}$, and $f_{z,n}$ are calculated according to \eqref{eq:sos_x_movement}, \eqref{eq:sos_y_movement}, and \eqref{eq:sos_z_movement} with $d=\Delta x=\Delta y=\Delta z=1$, respectively.

The iterative optimization is done by updating the $n$-th root frequency while keeping all other $N-1$ frequencies fixed. Then, the \ac{ASE} is calculated. If it improves, the update is applied, otherwise it is discarded and the previous value of $f_n$ is used. The update is calculated by
\begin{equation}\label{eq:sos_search}
    f_n = \frac{1}{d_S} \arg \min_{f} \sum_{s=1}^S
    \left\{ \rho(d_s) - \hat{\rho}(d_s) - \frac{1}{N} \cos
    \left( \frac{2\pi}{d_S} f d_s \right)\right\} ^2\text{,}
\end{equation}
where $\rho(d_s)$ is the desired \ac{ACF} and $\hat{\rho}(d_s)$ is the approximate \ac{ACF} constructed from all $N-1$ components from the previous iteration. The amplitudes of all sinusoids are set to $a_n^2 = \frac{2}{N}$ and standard numerical methods can be applied to find the values $f$. However, \eqref{eq:sos_search} can only be used to approximate a \ac{1-D} random process such as \eqref{eq:sos_1D_process}. For a \ac{3-D} random process \eqref{eq:sos_3D_process} it is necessary to estimate the three sinusoid components $f_{x,n}$, $f_{y,n}$, and $f_{z,n}$. This can be done by performing the estimation along the axes of the coordinate system. For example, combining \eqref{eq:sos_x_movement} and \eqref{eq:sos_approximate_acf} while setting $\Delta y = \Delta z = 0$ leads to a  directional \acp{ACF} in $x$-direction
\begin{equation}
    \hat{\rho}(\Delta x) = \frac{1}{N}\sum_{n=1}^N \cos ( 2\pi d_s \cdot \underbrace{f_n \cos \phi_n  \cos \theta_n}_{=f_{x,n}}  ) \text{.}
\end{equation}
This function is used instead of $\hat{\rho}(d_s)$ in \eqref{eq:sos_search} to get an update of the $n$-th root frequency $f_{n}$. Due to the linear dependency, $f_{x,n}$, $f_{y,n}$, and $f_{z,n}$ can be calculated from \eqref{eq:sos_x_movement}, \eqref{eq:sos_y_movement}, and \eqref{eq:sos_z_movement}. The same approach can be used to perform the estimation in $y$ or $z$-direction, where
\begin{eqnarray}
  \hat{\rho}(\Delta y) &=& \frac{1}{N}\sum_{n=1}^N \cos ( 2\pi d_s \cdot \underbrace{f_n \sin \phi_n \cos \theta_n}_{=f_{y,n}}  )\text{,} \\
  \hat{\rho}(\Delta z) &=& \frac{1}{N}\sum_{n=1}^N \cos ( 2\pi d_s \cdot \underbrace{f_n \sin \theta_n}_{=f_{z,n}}  ).
\end{eqnarray}
Depending on which of the $N$ sinusoid frequencies is estimated, the estimation direction is chosen according to
\begin{equation}
    \hat{\rho}(d_s) =
    \left\{
      \begin{array}{ll}
        \hat{\rho}(\Delta x), & \hbox{for } \Delta x \geq \Delta y \hbox{ and } \Delta x \geq \Delta z \hbox{;} \\
        \hat{\rho}(\Delta y), & \hbox{for } \Delta y > \Delta x \hbox{ and } \Delta y \geq \Delta z \hbox{;}\\
        \hat{\rho}(\Delta z), & \hbox{for } \Delta z > \Delta x \hbox{ and } \Delta z > \Delta y \hbox{,}
      \end{array}
    \right.
\end{equation}
where $\Delta x$, $\Delta y$, and $\Delta z$ are calculated according to \eqref{eq:sos_x_movement}, \eqref{eq:sos_y_movement}, and \eqref{eq:sos_z_movement} with $d=1$, respectively.

Cai and Giannakis \cite{Cai2003} introduced the \ac{ASE} as a performance measure of the approximation. It is defined as the average squared error between the desired \ac{ACF} $\rho(\mathbf{d})$ and the approximate \ac{ACF} $\hat{\rho}(\Delta x,\Delta y,\Delta z)$. Here it is calculated as
\begin{equation}\label{eq:sos_ASE_formula}
    \mathrm{ASE} = \frac{1}{S T}\sum_{t=1}^T\sum_{s=1}^S \left\{ \rho(d_s) - \frac{1}{N}\sum_{n=1}^N \cos \left( \frac{2\pi}{d_S}  f_{t,n}  d_s{} \right) \right\}^2 \text{.}
\end{equation}
Since the approximate \ac{ACF} depends on the direction, the \ac{ASE} calculation must take the directivity into account. Hence, the evaluation is done for $t = 1\ldots T$ \emph{test} directions $(\theta_t,\phi_t)$. The corresponding \emph{test} frequencies $f_{t,n}$ are
\begin{equation}
    f_{t,n} =  \left( f_{x,n} \cos \phi_t + f_{y,n} \sin \phi_t \right) \cos \theta_t + f_{z,n} \sin \theta_t\text{.}
\end{equation}
The \ac{ASE} is used as a cost function for the \emph{iterative refinement} of the sinusoid frequencies. If the \ac{ASE} improves for a newly estimated frequency, the sinusoid frequencies are replaced by the newly estimated ones, otherwise the new values are discarded. The iteration stops when no further improvement can be achieved for any of the $N$ frequencies.

The output of the approximation method are the sinusoid frequencies that can be used in \eqref{eq:sos_3D_process} to generate spatially correlated Normal-distributed random numbers with an arbitrary \ac{ACF}. It is possible to adjust the decorrelation distance $d_\lambda$ and the distribution function of the random process without having to recalculate the sinusoid frequencies. Doubling the distances in \eqref{eq:sos_dist_vec} is equal to halving the frequencies in \eqref{eq:sos_3D_process}. For example, if the approximation was done for $d_\lambda= 10$~m, but correlated random variables are needed for 20~m decorrelation distance, then simply dividing the frequencies by 2 creates the correct results. Uniform distribution can be achieved by a transformation from Normal to Uniform samples\footnote{Given a standard Normal distributed random variable $k \sim \mathcal{N}(0,1)$, then remapping the probability density to $u \sim \mathcal{U}(0,1)$ is done using the complementary error function as $u = 0.5\cdot\mathrm{erfc}( -k / \sqrt{2} )$.}. In the next section, it is shown how the \ac{SOS} method can be used to generate correlated random numbers for \ac{P2P} links where both ends of the communication channel are mobile.

\section{Device-to-Device Extension}

It was found in \cite{Wang2005b} that the \ac{MT} movement at each end of the \ac{P2P} link has an independent and equal effect on the correlation coefficient and that the \ac{JCF} can be decomposed into two independent \acp{ACF}
\begin{multline}\label{eq:independent_jcf}
    \rho(\Delta x_t,\Delta y_t,\Delta z_t,\Delta x_r,\Delta y_r,\Delta z_r ) = \\
        \rho(\Delta x_t,\Delta y_t,\Delta z_t) \cdot \rho(\Delta x_r,\Delta y_r,\Delta z_r).
\end{multline}
The locations of the transmitting and receiving terminal are given in \ac{3-D} Cartesian coordinates as $(x_t,y_t,z_t)$ and $(x_r,y_r,z_r)$, respectively. \cite{Wang2005b} proposes to approximate a combined \ac{ACF}. However, with six dimensions this results in a prohibitively large number of sinusoid coefficients and computing time. A simpler approach is to combine two independent Gaussian random processes, one for the transmitter and one for the receiver into
\begin{equation}\label{eq:sos_6D_process}
    k(x_t,y_t,z_t,x_r,y_r,z_r) = \frac{k_t(x_t,y_t,z_t) + k_r(x_r,y_r,z_r)}{\sqrt{2}}.
\end{equation}
The approximated six-dimensional Gaussian process is
\begin{multline}\label{eq:sos_approx_6D_process}
    \hat{k}(x_t,y_t,z_t,x_r,y_r,z_r) = \\
        \sum_{n=1}^{2N} \frac{a_n}{\sqrt{2}} \cos\left\{ 2\pi \cdot \mathbf{f}_n^T \cdot
        \left[  x_t \enspace y_t \enspace z_t \enspace x_r \enspace y_r \enspace z_r \right]^T + \psi_n
        \right\}\text{,}
\end{multline}
where the sinusoid frequencies are obtained independently for the transmitting and receiving side.
\begin{equation}
    \mathbf{f}_n =
    \left\{
      \begin{array}{ll}
        \left[
            \begin{array}{cccccc}
              f_{x_t,n} & f_{y_t,n} & f_{z_t,n} & 0 & 0 & 0
            \end{array}
         \right]^T, & \hbox{for } n \leq N \hbox{;} \\
        \left[
            \begin{array}{cccccc}
              0 & 0 & 0 & f_{x_r,n} & f_{y_r,n} & f_{z_r,n}
            \end{array}
        \right]^T, & \hbox{for } n > N \hbox{.}
      \end{array}
    \right.
\end{equation}
The scaling by $\sqrt{2}$ in \eqref{eq:sos_6D_process} accounts for doubling the number of \ac{SOS} components in \eqref{eq:sos_approx_6D_process}.

\section{Numeric Results}

\begin{figure}[t]
    \centering
    \includegraphics[width=69mm]{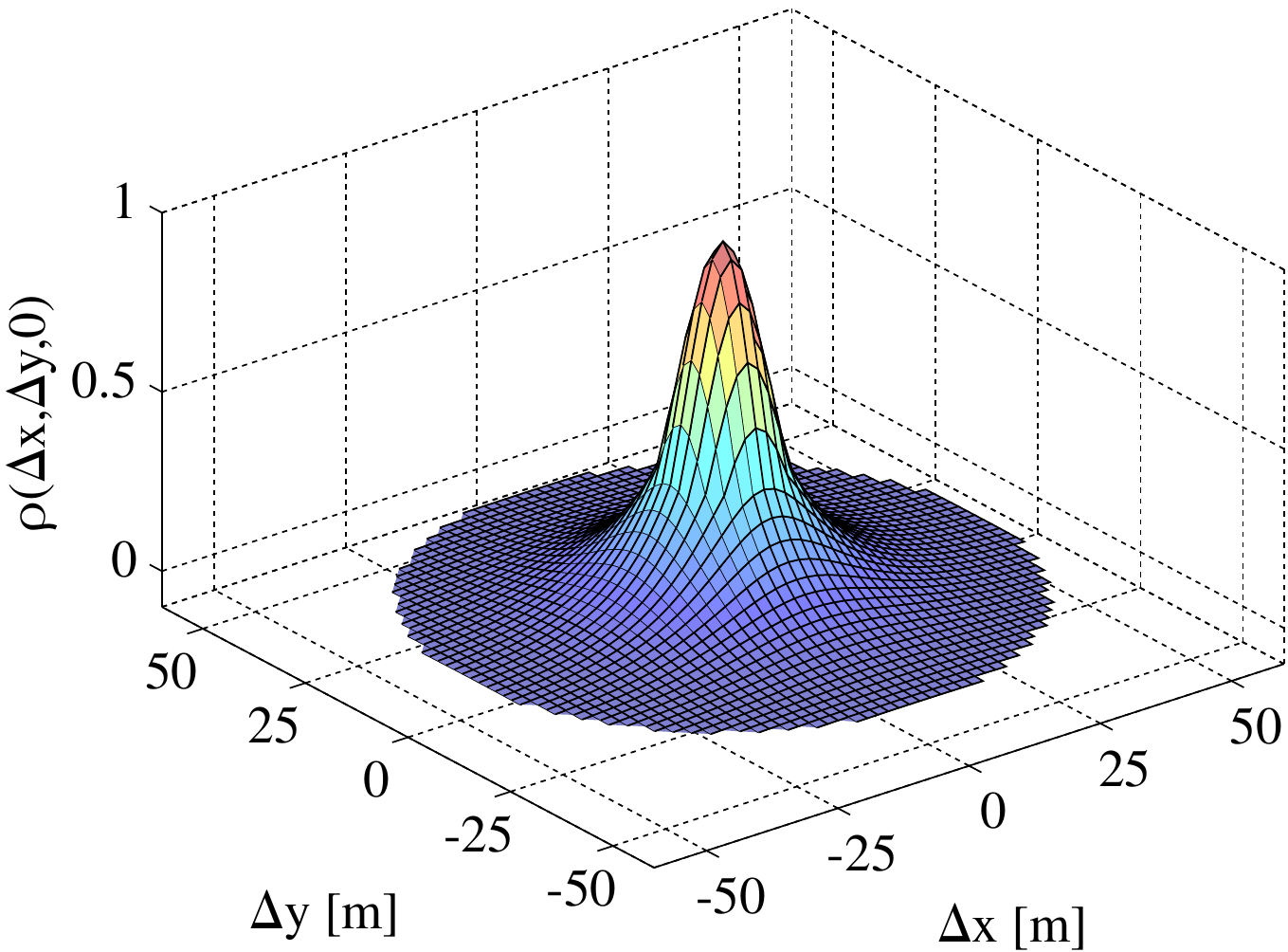}
    \includegraphics[width=69mm]{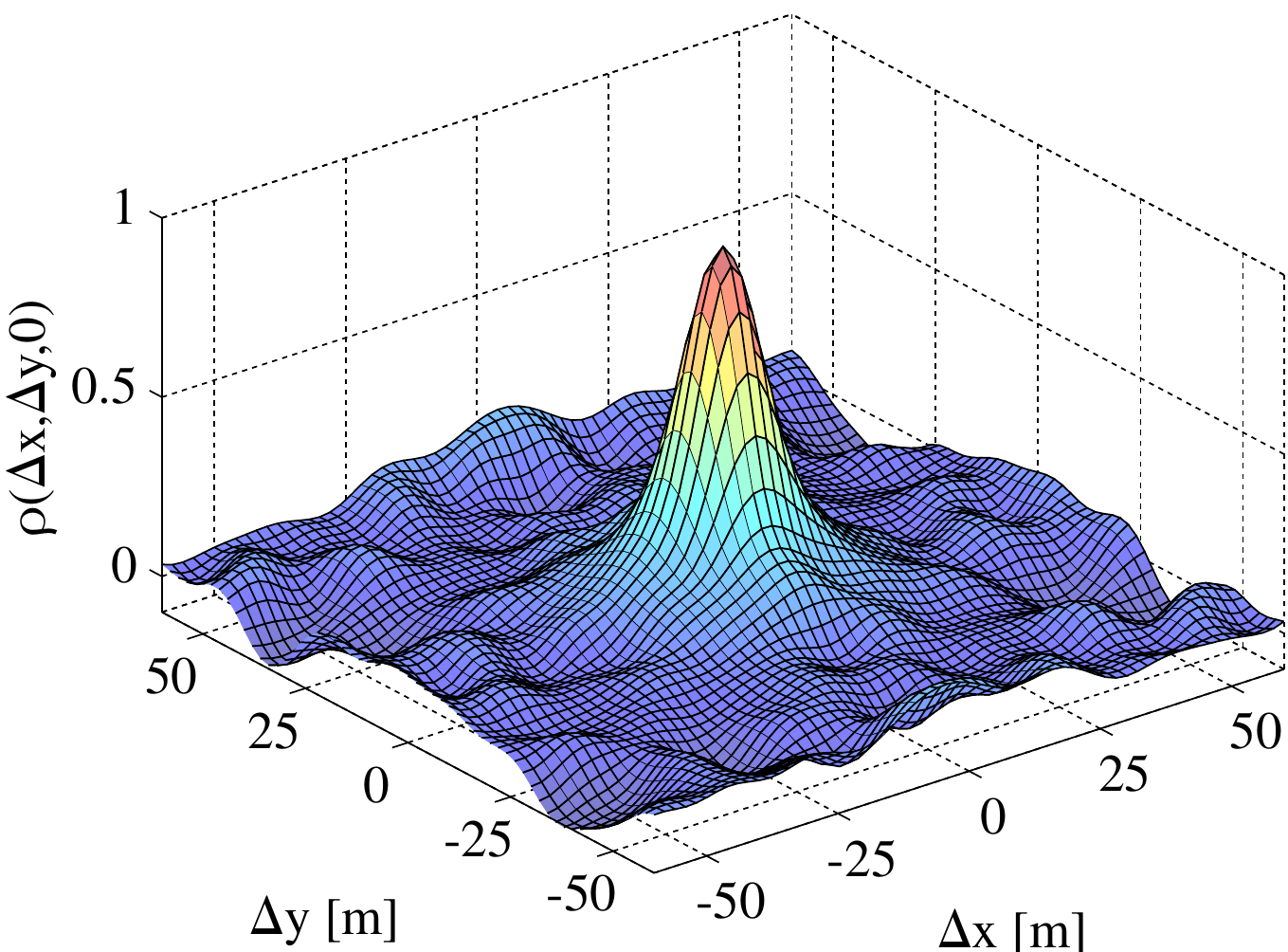}
    \vspace{-2mm}
    \caption{Desired 2-D ACF on the $x-y$ plane (top) and approximated ACF using 300 sinusoids (bottom)}
    \label{fig:acf_2d_desired}
\end{figure}

\begin{figure}[t]
    \centering
    \vspace{-2mm}
    \includegraphics[width=72mm]{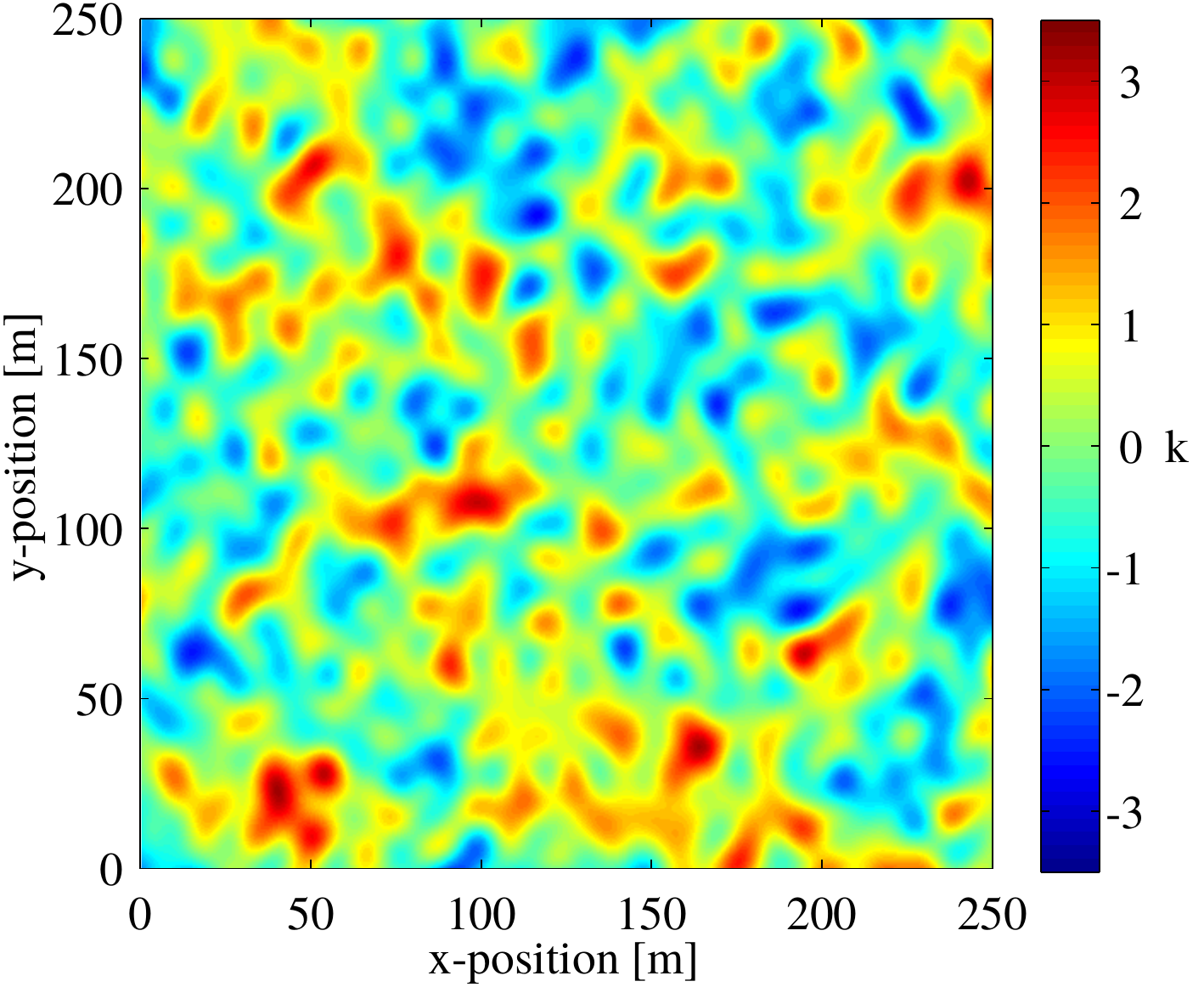}
    \vspace{-2mm}
    \caption{Example of a spatially correlated 2-D process}
    \label{fig:sample_2d_process}
\end{figure}

In the following, we present numerical results for the proposed estimation method. To demonstrate the approximation of an arbitrary \acp{ACF}, we study the performance of the model for a combination of a Gaussian and an exponential \ac{ACF} which has the same decorrelation distance as \eqref{eq:exponential_corr_decay_map}
with closely spaced samples showing a higher correlation.
\begin{equation}
    \rho(d) =
    \left\{
      \begin{array}{ll}
        \exp\left(-\frac{d^2}{d_\lambda^2}\right), & \hbox{for $d < d_\lambda$;} \\
        \exp\left(-\frac{d}{d_\lambda}\right), & \hbox{for $d \geq d_\lambda$}
      \end{array}
    \right.
\end{equation}
Results for the exponential \ac{ACF} are presented at the end of this section. The decorrelation distance $d_\lambda$ was set to 10 meters and the \ac{ACF} was sampled in intervals of 0.25 meters ranging from 0 to 49.75 meters. Hence, $\rho(d_s)$ contains 200 values. A \ac{2-D} plot of this function is shown in Fig.~\ref{fig:acf_2d_desired} (top). This function was approximated using the method outlined in section \ref{sec:sos_arbitrary_approx} with a varying number of sinusoid coefficients. The number of test directions for the iterative search was set to 28 and the entire estimation process was repeated 5000 times with different initial values for the directions $(\theta_n,\phi_n)$ and the initial root-frequencies. Fig.~\ref{fig:acf_2d_desired} (bottom) shows the best result of an approximation using 300 sinusoids. An example of the \ac{2-D} random process resulting from this approximation is shown in Fig.~\ref{fig:sample_2d_process}. The output was generated by defining $x,y$-positions on a 250 $\times$ 250~m grid with 0.5~m spacing. The $z$-position was set to 0~m and no \ac{D2D} extension was used.

\paragraph{ACF approximation accuracy}

\begin{figure}[h]
    \centering
    \includegraphics[width=69mm]{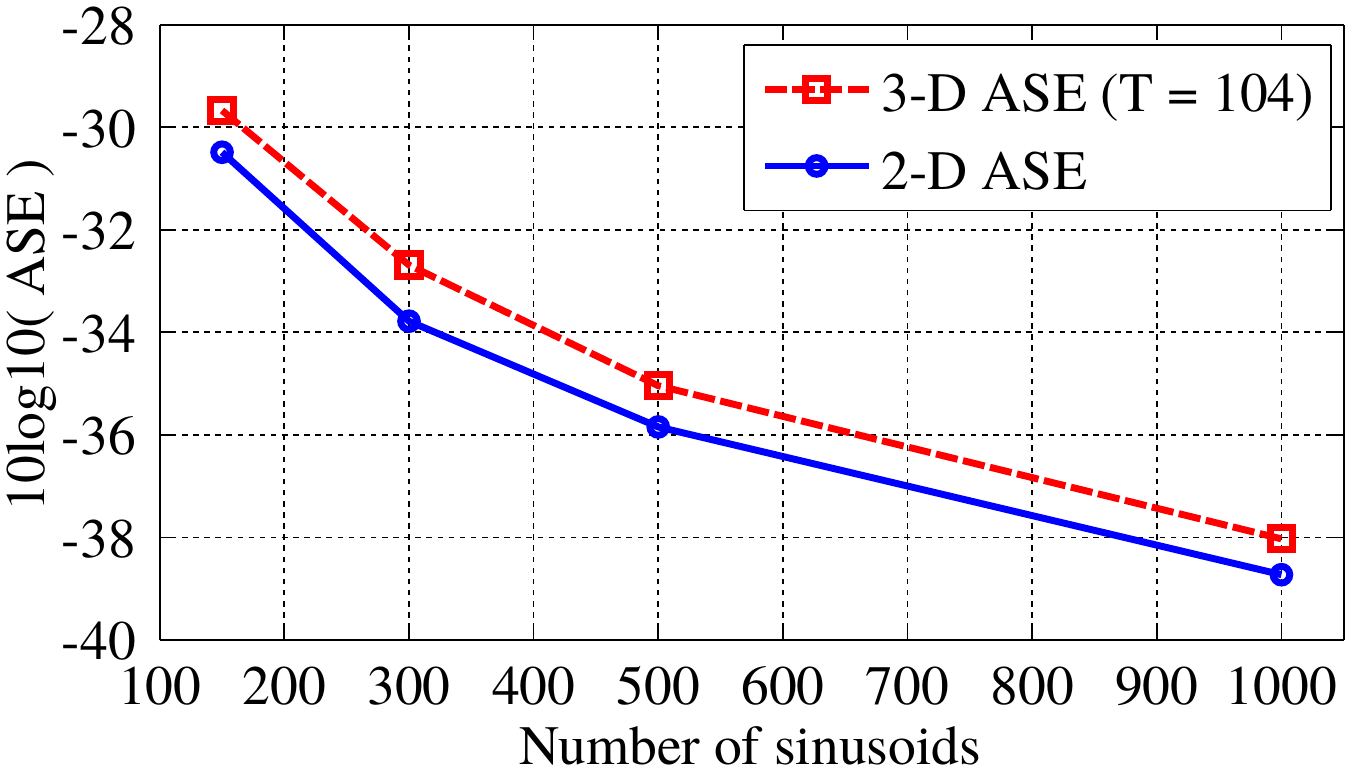}
    \vspace{-2mm}
    \caption{Achievable ASE vs. number of sinusoids for the combined ACF}
    \label{fig:sos_performance}
\end{figure}

The achievable \ac{ASE} as a function of the number of sinusoids is shown in Figure~\ref{fig:sos_performance} (lower values are better). The red dashed curve shows the \ac{ASE} calculation using \eqref{eq:sos_ASE_formula} whereas the blue, solid line compares the \ac{2-D} \acp{ACF} on the $x-y$, $x-z$, and $y-z$ plane with the one shown in Fig.~\ref{fig:acf_2d_desired} (taking only the disk-shaped area covered by the \ac{ACF} into account). Both methods estimate similar values for the \ac{ASE}. The curves indicate that when doubling the number of coefficients, the \ac{ASE} improves by roughly 3~dB. However, this also comes at the cost of doubling the computation time for the generation of the random numbers.

\paragraph{Random number generation}

\begin{figure}[b]
    \centering
    \includegraphics[width=70mm]{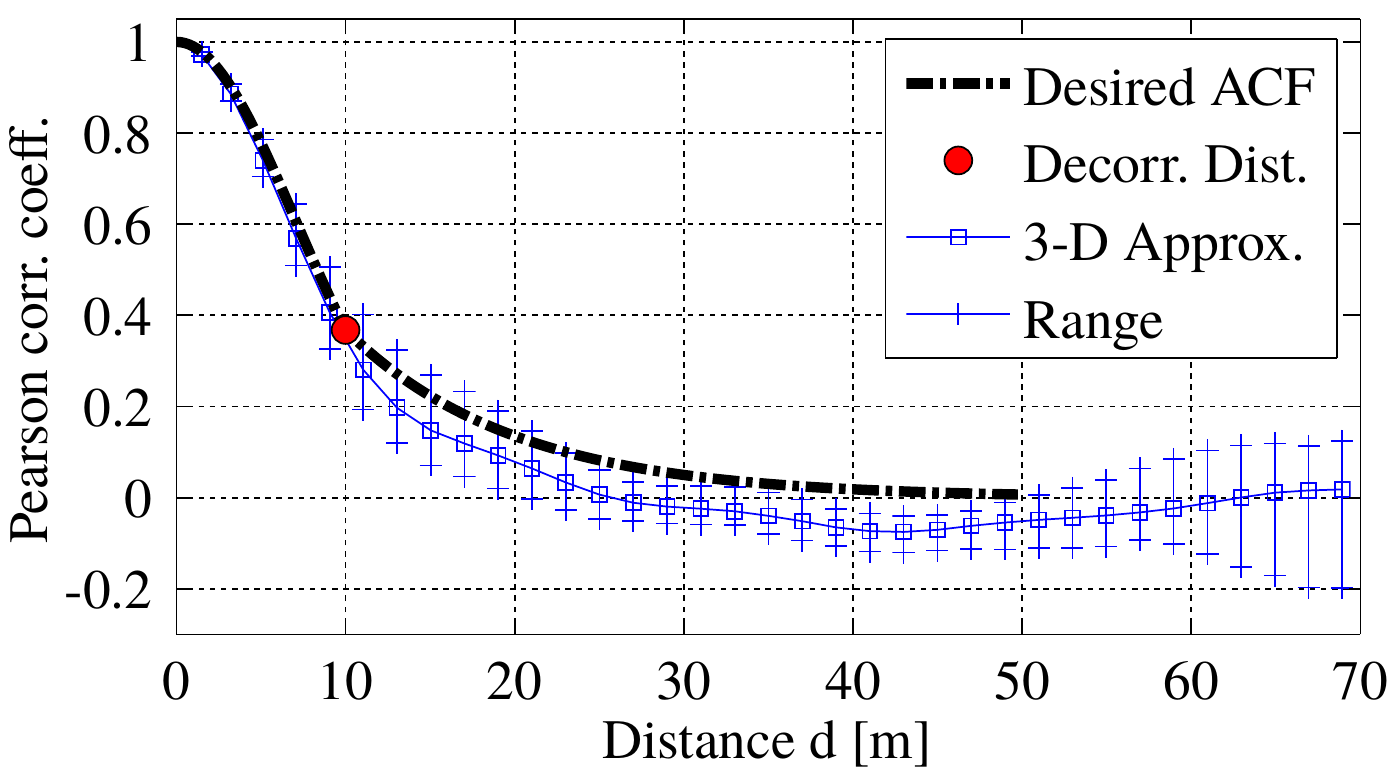}\vspace{2mm}
    \includegraphics[width=70mm]{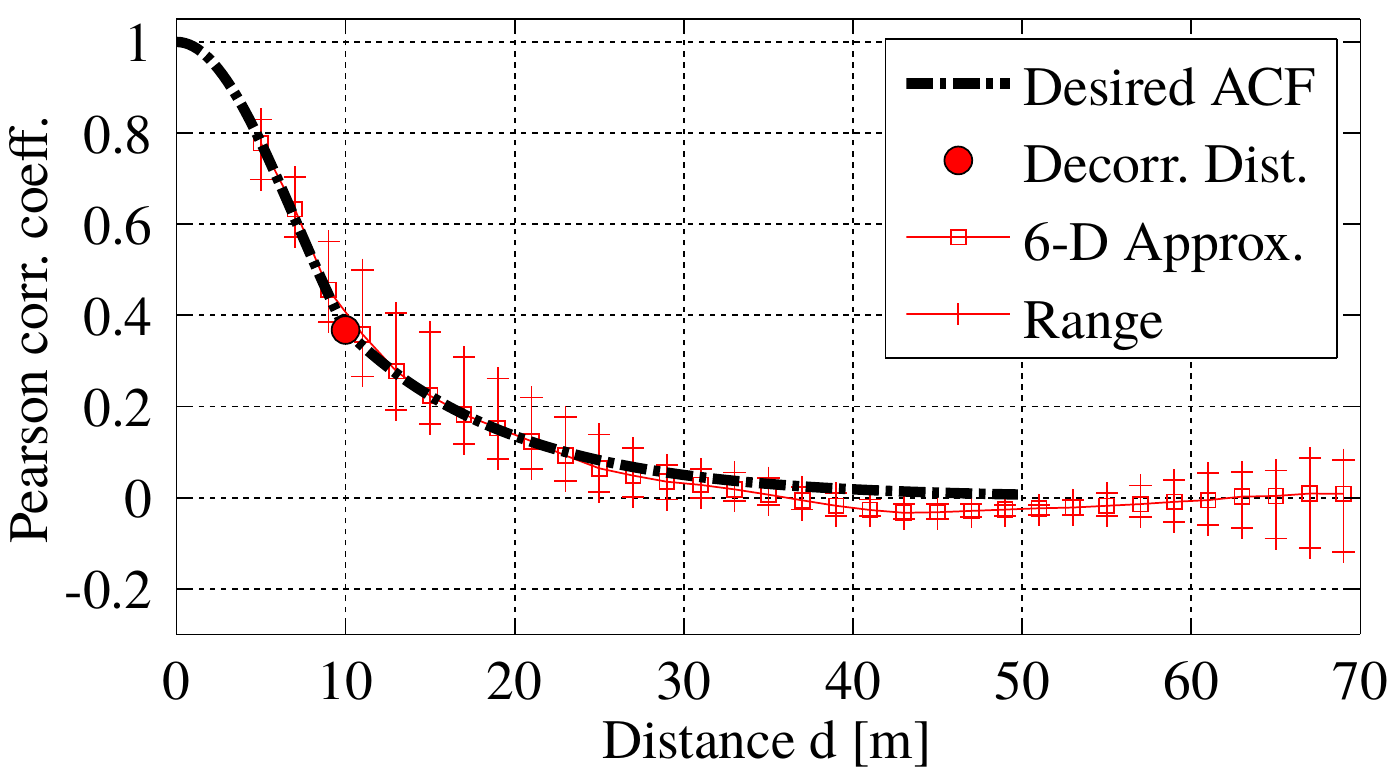}
    \vspace{-2mm}
    \caption{Estimated ACF from spatially correlated random values}
    \label{fig:acf_1d}
\end{figure}

Figure~\ref{fig:acf_1d} compares the desired \ac{ACF} (thick dashed line) with the output of the random number generator using 300 sinusoids. For the upper plot, 6000 positions were randomly chosen within a cube of 56~m edge length. Then, spatially correlated Normal distributed random numbers were generated for each position using \eqref{eq:sos_3D_process}. The distance between each pair of positions was calculated and pairs with similar distance were grouped, i.e.\ positions with a distance between 0 and 2 meters of each other belong to group 1, positions with a distance between 2 and 4 meters belong to group 2, and so on. The Pearson correlation coefficient was calculated for the samples within each group. The entire process was repeated 10 times to obtain an average \ac{ACF} (blue squares) and the spread for different initializations (bars showing the minimum and maximum values of the 10 initializations). For the lower part of the figure, 3000 positions were randomly chosen for the transmitter and 3000 were chosen for the receiver. Then, spatially correlated values were generated using \eqref{eq:sos_6D_process}. The grouping was done for the transmitter positions under the constraint that the receiver does not move faster than the transmitter. For example, group 3 contains all pairs where the transmitters have a distance between 4 and 6 meters and the receivers have a distance between 0 and 6 meters of each other. Both curves indicate that the given \ac{ACF} can be approximated with good accuracy.

\begin{figure}[h]
    \centering
    \includegraphics[width=69mm]{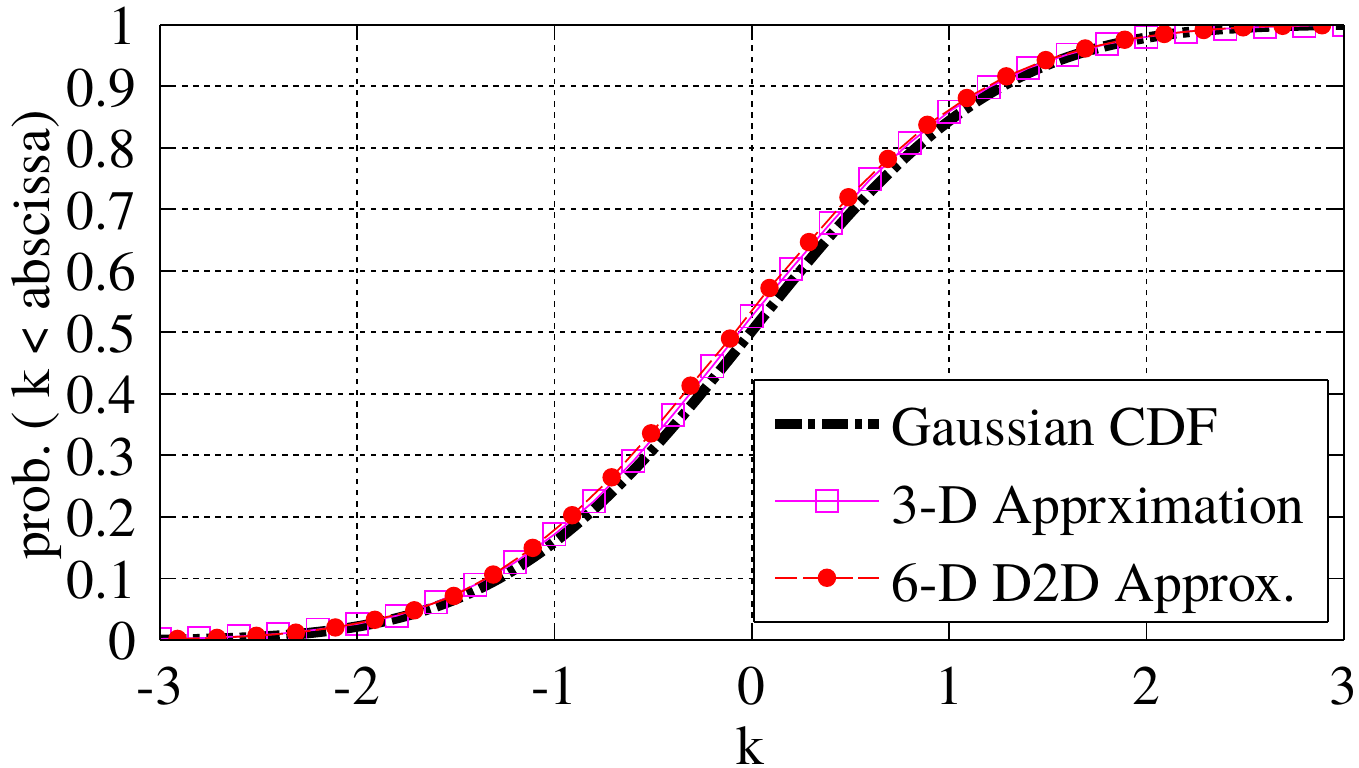}
    \vspace{-2mm}
    \caption{CDF of the output values vs. the Gaussian CDF}
    \label{fig:sos_cdf}
\end{figure}

Figure~\ref{fig:sos_cdf} shows that the \ac{CDF} of the generated random numbers is close to Gaussian density. For this, $10^4$ random positions were generated in a cube of $1000 \times 1000 \times 50$~m edge length and spatially correlated random numbers were generated using \eqref{eq:sos_3D_process} and \eqref{eq:sos_6D_process}. There is no notable difference between the distribution of the generated values and the Gaussian \ac{CDF}.

\paragraph{Complexity considerations}

As mentioned in the introduction, the state-of-the-art alternative for the generation of correlated random numbers is by filtering white noise \cite{Bakowski2011}. Here, the filter approach is compared with the \ac{SOS} model in terms of memory requirements, initialization time, time to generate the output values, and the restrictions on the \ac{MT} placement. The results are summarized in Table~\ref{tab:sos_filter_performace_comparison}.

\begin{table}[h]
    \vspace{-2mm}
    \setlength{\tabcolsep}{0pt}
    \centering
    \caption{Performance comparison of filter-based and SOS model}
    \label{tab:sos_filter_performace_comparison}
\begin{tabular}{L{27mm} | C{29mm} | C{29mm}  }
    & Filter approach & Sum-of-sinusoids \\
    \hline
    Required memory         & $N^D$ elements  & $(D+1)\cdot N$ elements  \\
    Initialization          & $\approx D \cdot S \cdot N^D$ op. & $\approx N$ operations \\
    Output generation       & $2D^2 + 5D$ op. & $\approx N\cdot (2D+15)$ op.\\
    MT placement            & inside map area & no restriction \\
\end{tabular}
\end{table}

For the filter approach, the parameter $N$ in Tab.~\ref{tab:sos_filter_performace_comparison} refers to the \emph{edge length} of the sampled area and $D$ refers to the number of dimensions. The sample resolution needs to be high enough to captures the variation in the random process. With $d_\lambda=10$~m, we assume that a resolution of 2.5~m is sufficient. With $N=300$ and $D=2$, the filtered map covers an area of $750 \times 750$ meters and requires 360 kilobytes of memory\footnote{Assuming single-precision floating point accuracy.}. Without increasing $N$ or lowering the resolution, all \acp{MT} must be located within this area. In comparison, the \ac{SOS} method requires only 3.6 kilobytes of memory for the sinusoid coefficients and there are no restrictions on the \ac{MT} positions.

Initializing the random number generator means that a new map must be generated, i.e.\ for the filter approach, generating $N^D$ random numbers and processing each number with $D$ finite impulse response filters\footnote{The filter order $S$ corresponds to the length of the sampled \ac{ACF}.} of order $S$. For the \ac{SOS} method, only the $N$ phases $\psi_n$ must be randomly initialized in \eqref{eq:sos_3D_process}. The sinusoid frequencies can be precalculated and stored in a table. This makes the \ac{SOS} method very efficient when many random variables are needed, such as for the spacial consistency model from 3GPP \cite{3gpp_tr_38901_v1410}. The downside of the \ac{SOS} method resides in the commutation time required to generate the correlated output values. Calculating the cosine in \eqref{eq:sos_3D_process} is computationally expensive\footnote{Calculating the cosine requires approximately 15 floating point operations, depending on the implementation and the value of the argument.} and the sum over all $N$ sinusoids must be calculated for each output value. With $N=300$ and $D=2$, approximately 5700 operations are needed to generate one output value. For the filter approach, the map only needs to be interpolated to obtain the output value at the \ac{MT} position. This requires 18 operations using bilinear interpolation. Hence, the filter approach might be preferred when the output value generation is the dominating factor in the overall computing time.

\paragraph{Exponential ACF approximation}

\begin{figure}[h]
    \centering
    \includegraphics[width=69mm]{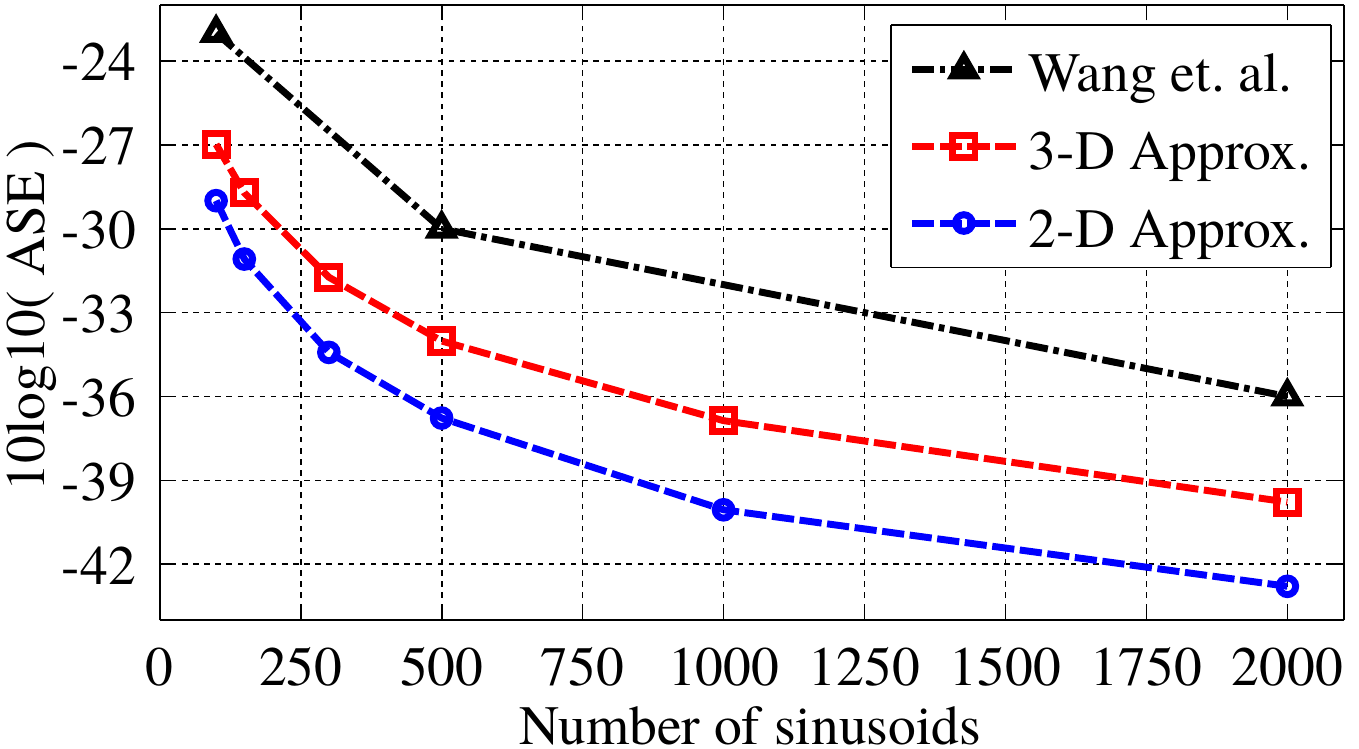}
    \vspace{-2mm}
    \caption{Achievable ASE vs. number of sinusoids for the exponential ACF}
    \label{fig:sos_performance_Exp}
\end{figure}

The achievable \ac{ASE} performance is compared with the results from Wang et.\ al. \cite{Wang2005b}, where an exponential \ac{ACF} was approximated by different methods using 100, 500, and 2000 sinusoids. With the Monte Carlo method, \cite{Wang2005b} reported an \ac{ASE} for a \ac{2-D} approximation of $-23$~dB, $-30$~dB, and $-36$~dB, respectively. The other proposed methods performed similarly. Our iterative approximation method achieved $-29$~dB, $-36.8$~dB, and $-42.7$~dB for the \ac{2-D} approximation of the exponential \ac{ACF} using the same number of sinusoids, respectively. This is an average improvement of $6.5$~dB. Results are shown in Fig.~\ref{fig:sos_performance_Exp}. The \ac{3-D} approximation has an average performance loss of 2.7~dB compared to \ac{2-D}\footnote{The performance loss can be explained by the additional degree of freedom added by the mobility in $z$-direction in \eqref{eq:sos_z_movement}, where the scaling of the resolution in the $x,y$-direction follows from $\int_{-\pi/2}^{\pi/2}\cos^2\theta d\theta / \int_{-\pi/2}^{\pi/2} d\theta = \frac{1}{2}$. }. However, this is still a significant improvement compared to \cite{Wang2005b}.

\newpage
\section{Conclusions}

The proposed iterative approximation method can obtain the sinusoid coefficients for arbitrary \acXp{ACF} with better accuracy compared to previous results. It is possible to change the decorrelation distance and the distribution of the spatially correlated random process without recalculating the sinusoid frequencies. They can be precomputed and stored in a table for fast access. Therefore, generating spatially correlated random numbers, even for a large number of positions, can be done by simply calculating a weighted sum. The performance scales linearly with the number of sinusoids and the number of positions. Furthermore, memory requirements are negligible. This provides a very efficient implementation of the 3GPP proposal for spatial consistency. The scheme can be used with minimal adjustments to support a system where both ends of a link are mobile.

\section*{Acknowledgement}

The authors thank the Celtic Office and national funding authorities BMBF in Germany, Business Finland, and MINETAD in Spain for supporting this research and development through the ReICOvAir project. The project benefits also from the valuable technical contributions from GHMT AG, CETECOM GmbH, and Qosmotec GmbH in Germany; Trimek S.A. and SQS S.A. in Spain; Verkotan Ltd., Kaltio Technologies, Sapotech, and the Centre for Wireless Communications at the University of Oulu in Finland.

\vfill
\bibliographystyle{IEEEtran}
\bibliography{sos}

\end{document}